\def\cm{{\rm\thinspace cm}}

\def\erg{{\rm\thinspace erg}}

\def\g{{\rm\thinspace g}}
\def\K{{\rm\thinspace K}}
\def\keV{{\rm\thinspace keV}}
\def\km{{\rm\thinspace km}}
\def\kpc{{\rm\thinspace kpc}}

\def\Mpc{{\rm\thinspace Mpc}}
\def\Msun{\hbox{$\rm\thinspace M_{\odot}$}}
\def\Zsun{\hbox{$\rm\thinspace Z_{\odot}$}}
\def\pc{{\rm\thinspace pc}}

\def\s{{\rm\thinspace s}}
\def\yr{{\rm\thinspace yr}}

\def\cmpssq{\hbox{$\cm\s^{-2}\,$}}

\def\pcmcu{\hbox{$\cm^{-3}\,$}}

\def\ergps{\hbox{$\erg\s^{-1}\,$}}

\def\gpcmsq{\hbox{$\g\cm^{-2}\,$}}

\def\kmps{\hbox{$\km\s^{-1}\,$}}
\def\kmpspmpc{\hbox{$\km\,\s^{-1}\Mpc^{-1}\,$}}

\def\Msunppc{\hbox{$\Msun\pc^{-1}\,$}}

\def\Msunpyr{\hbox{$\Msun\yr^{-1}\,$}}

\def\pcmsq{\hbox{$\cm^{-2}\,$}}
\def\pcmK{\hbox{$\cm^{-3}\K$}}

\documentclass[usegraphicx]{mn2e}
\begin{document}

\title[Filaments and Swirl in the Centaurus cluster]{HST imaging of
  the dusty filaments and nucleus swirl in NGC4696 at the centre of
  the Centaurus Cluster
  } \author[A.C. Fabian et al] {\parbox[]{6.5in}{{
      A.C.~Fabian$^1\thanks{E-mail: acf@ast.cam.ac.uk}$,
      S.A. Walker$^1$, H.R. Russell$^1$, C. Pinto$^1$,
      R.E.A. Canning$^2$, P. Salome$^3$, J.S. Sanders$^4$,
      G.B. Taylor$^5$, E.G. Zweibel$^{6,7}$, C.J. Conselice$^8$,
      F. Combes$^3$, C.S. Crawford$^1$, G.J. Ferland$^9$,
      J.S. Gallagher III$^6$, N.A. Hatch$^7$, R.M. Johnstone$^1$ and
      CS. Reynolds$^{10}$
    }\\
    \footnotesize $^1$ Institute of Astronomy, Madingley Road,
    Cambridge CB3 0HA\\
    $^2$ KIPAC, Stanford University, 452, Lomita Mall, Stanford, CA
    94305-4085, USA\\
    $^3$ LERMA, Obs, de Paris, UMR8112, 61 Av. de
    l'Observatoire, Paris, France\\
    $^4$ Max Planck Institut fur Extraterrestrische Physik,
    Giessenbachstrasse 1, 85748 Garching,
    Germany\\
    $^5$ Dept. of Physics and Astronomy, University of New
    Mexico, Alburquerque, NM 87131, USA\\
    $^6$ Dept. of Astronomy,
    University of Wisconsin, WI 53706, USA\\
    $^7$ Dept. of Physics,
    University of Wisconsin, WI 53706, USA\\
    $^8$ School of Physics \&
    Astronomy, University of Nottingham, Nottingham, NG7 2RD\\
    $^9$ Dept. of Physics and Astronomy, University of Kentucky,
    Lexington
    KY 40506, USA \\
    $^{10}$ Dept. of Astronomy, University of Maryland, College Park,
    MD 20712-2421, USA\\
  }}

\maketitle
  
\begin{abstract}
  Narrow-band HST imaging has resolved the detailed internal structure
  of the 10 kpc diameter H$\alpha$+[NII] emission line nebulosity in NGC4696,
  the central galaxy in the nearby Centaurus cluster, showing that the
  dusty, molecular, filaments have a width of about 60pc. Optical
  morphology and velocity measurements indicate that the filaments are
  dragged out by the bubbling action of the radio source as part of
  the AGN feedback cycle. Using the drag force we find that the
  magnetic field in the filaments is in approximate pressure
  equipartition with the hot gas. The filamentary nature of the cold
  gas continues inward, swirling around and within the Bondi accretion
  radius of the central black hole, revealing the magnetic nature of
  the gas flows in massive elliptical galaxies. HST imaging resolves
  the magnetic, dusty, molecular filaments at the centre of the
  Centaurus cluster to a swirl around and within the Bondi radius.
\end{abstract}


\section{Introduction}
The Brightest Cluster Galaxy (BCG), NGC4696, in the nearby Centaurus
cluster, Abell 3526, has a system of dusty H$\alpha$ filaments (Fabian
et al. 1982; Sparks et al. 1989, Crawford et al 2005) connecting the
hot intracluster medium (ICM) with its Active Galactic Nucleus
(AGN). The low star formation rate ($0.1\Msunpyr$) means that energy
feedback into the ICM via the AGN-powered radio source has stifled
almost all complete radiative cooling in the hot gas for billions of
years, despite a central cooling time (Sanders et al. 2008, 2016) of
only $\sim3\times10^7\yr$. The details of how feedback is so
successful and the operation of key ingredients such as the heating of
the hot gas, the behaviour of the cold gas, the lack of star
formation, and the fuelling of the central back hole are unclear
(Fabian 2012). The filaments provide diagnostic information on the
distribution and morphology of cold gas and its velocity field
(Canning et al 2010) in the region where much of the central action
takes place. The Centaurus system is important since it is the nearest
BCG with detected, extended, molecular gas. Here we present Hubble
Space Telescope (HST) H$\alpha$ images of the filaments which show
that they are likely to be magnetically-supported in the hot gas,
similar to the situation for the spectacular filament system around
NGC1275 in the more distant Perseus cluster (Fabian et al 2008). The
images also reveal an intriguing swirl in the cold gas at the Bondi
radius of the black hole.

The temperature of the hot gas in the central 20 kpc radius of the
Centaurus cluster drops from 3 keV down to a mean of about
0.5keV. Chandra observations show that the gas is multiphase within
10kpc with a 0.5 keV component extending over the 5 kpc radius region
where the optical filaments are seen (Sanders et al 2016). XMM-Newton
Reflection Grating Spectrometer data reveal strong FeXVII emission
lines from the 0.5 keV gas indicating that any complete cooling flow,
which would be about $40 \Msunpyr$ if no heating took place, is
reduced to below $4\Msunpyr$ at temperatures below 0.4~keV (Sanders et
al 2008). The filaments consist of both an atomic component giving
rise to the H$\alpha$ emission, as well as [NII], [OII], OI, [SiII]
characteristic of lowly-ionized gas, significant masses of cold gas
indicated by [CII] emission (Mittal et al 2011) and a molecular
component is seen in H$_2$ rotational lines with Spitzer (Johnstone et
al 2007).

The low-ionization optical and near infrared emission spectrum of the
filaments strongly indicates that energetic particles are responsible
for heating and excitation of the cold gas (Ferland et al 2009). These
are either due to penetration of the cold gas by the surrounding hot
gas (Fabian et al 2011) or internal particles energised by magnetic
reconnection as the filaments move around (Churazov, Ruszkowski \&
Schekochihin 2013). If it is by penetration then the filaments are
growing at the expense of the hot gas at a few $\Msunpyr$. Unlike the
filaments in NGC1275, those in NGC4696 contain dust lanes, first
reported by Shobbrook (1963), which contrasts with the NGC1275
filaments for which no dust lanes have been detected. Dust has also
been detected in emission with Spitzer (Kaneda et al 2007) and
Herschel (Mittal et al 2011) with a total cold dust mass of
$1.6\times10^6 \Msun$. In comparison with the total molecular mass, the
dust-to- gas ratio appears to be normal. Comparison of the dust
extinction in the R and V bands shows (Sparks et al 1989) that to be
normal also.

The redshift of NGC\,4696 is 0.0104 and we adopt $H_0=71\kmpspmpc.$

\section{Filament Properties} 

10.5 HST orbits were obtained with Wide
Field Camera 3 (WFC3) using redshifted H$\alpha$ filter F665N. This
filter includes the H$\alpha$ and [NII] emission lines with [NII]
being the stronger component: for simplicity we refer to the combined
emission as H$\alpha$. The images were drizzled together and a
smoothed galaxy continuum model subtracted to give Fig. 1. The
filaments are resolved with a width of about 60 pc (Fig. 2). More
details of the reduction of the data are given in the Appendix.

\begin{figure}
  \centering
  \includegraphics[width=0.99\columnwidth,angle=0]{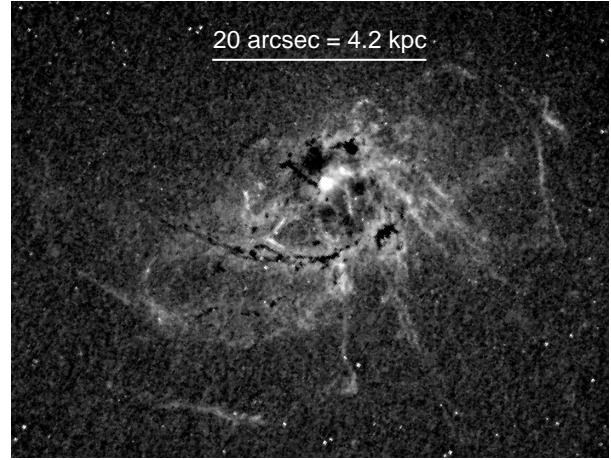}
  \caption{H$\alpha$+[NII] image of NGC4696.   }
\end{figure}

\begin{figure}
  \centering
  \includegraphics[width=0.99\columnwidth,angle=0]{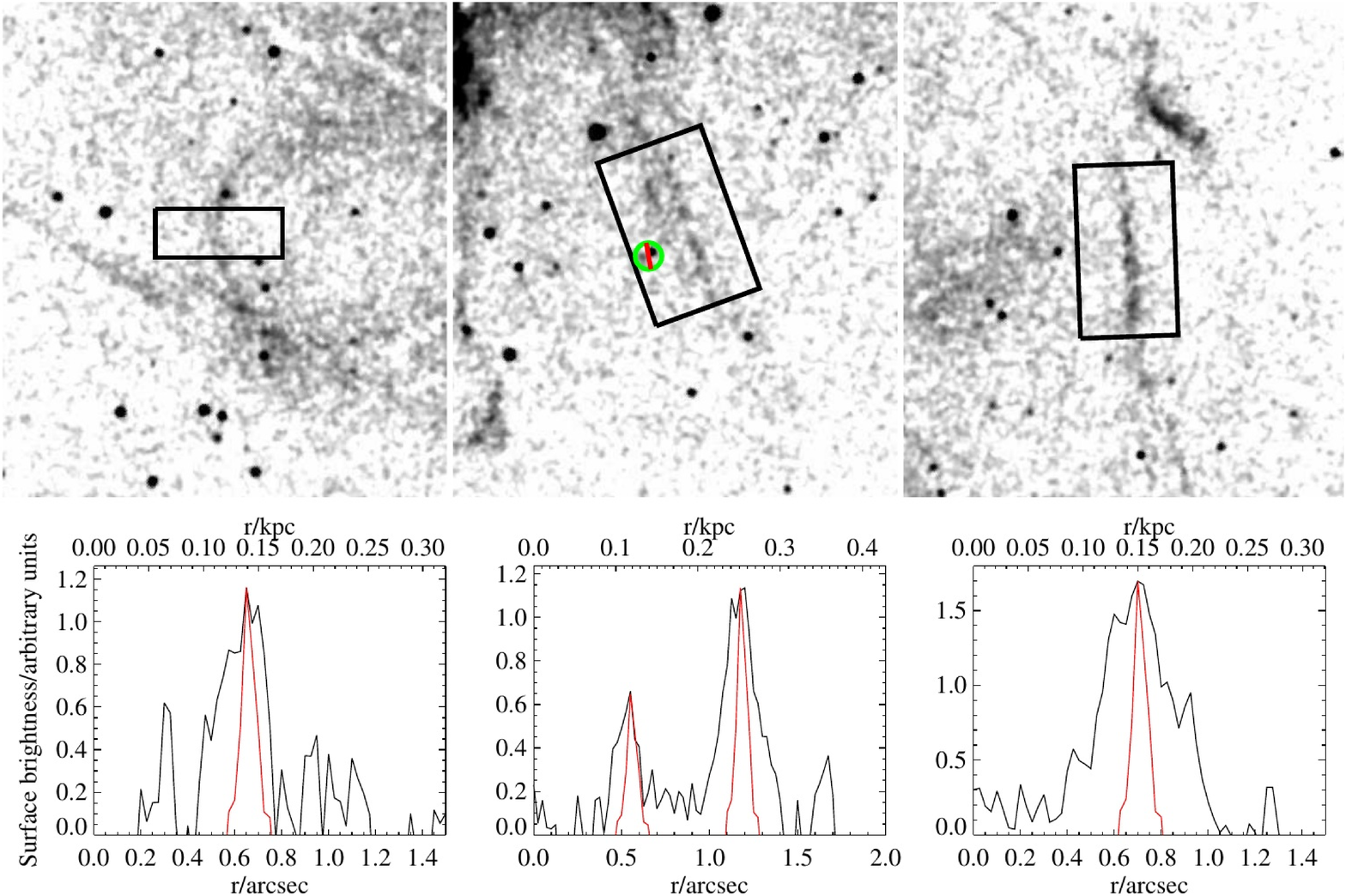}
  \caption{The top panels show three of the filaments in Centaurus. The
top left is part of the loop to the east, the top middle is a double
filament to the south, and the top right is a linear filament to the
west. Surface brightness profiles were taken across the width of these
filaments using the regions within the black boxes, and are shown in
the bottom panels respectively. These profiles are compared to the
profiles across nearby stars (red), showing that the filaments are
clearly resolved.   }
\end{figure}

\begin{figure}
  \centering
  \includegraphics[width=0.99\columnwidth,angle=0]{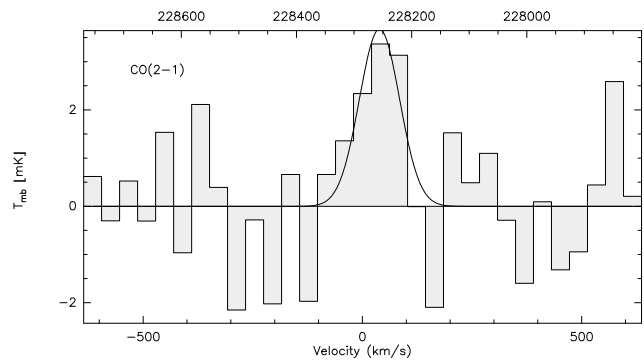}
  \caption{APEX spectrum showing 3.4$\sigma$ detection of CO(2-1) in NGC4696   }
\end{figure}

\begin{figure}
  \centering
  \includegraphics[width=0.99\columnwidth,angle=0]{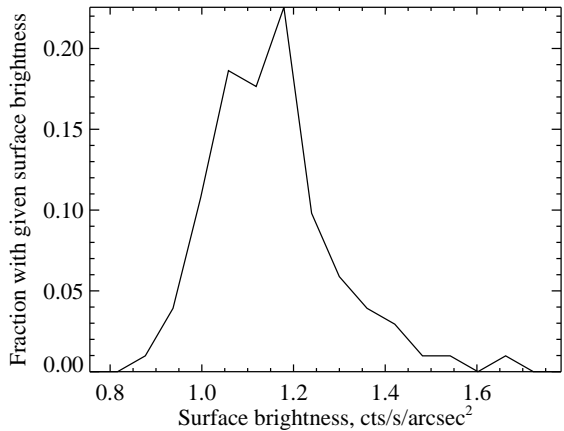}
  \caption{Histogram of the surface brightness along the outer H$\alpha$
filaments, demonstrating remarkable unformity.   }
\end{figure}

A $3.4\sigma$ detection of CO emission has been made using the Atacama
Pathfinder EXperiment (APEX, Fig. 3). Using a Milky Way conversion
factor (Bolatto et al 2013) we obtain a molecular mass of about
$10^8\Msun$. The ratio of total H$\alpha$ luminosity (Canning et al
2011) of $2.2 \times10^{40} \ergps$ to CO luminosity is similar to
that of other BCG systems (Edge et al 2001; Salom\'e \& Combes 2003).

Assuming that this ratio is constant over the NGC4696 filament system,
we find from the H$\alpha$ surface brightness that the linear mass
density of the filaments is about $1000\Msunppc$. The mean density of
a filament then corresponds to about 3 protons cm$^{-3}$ which is about 10
times denser than the surrounding hot gas. Of course the density of
the molecular gas within a filament must correspond to much higher
values: the thermal pressure is about $3\times 10^6 \pcmK$ so the CO
component is plausibly denser than $3\times10^4 \pcmcu$. The filaments are not
continuous cold structures but must contain smaller dense structures
or threads (Fabian et al 2008).

Inspection of the outer filaments suggests that they have similar
surface brightness. This is confirmed using half-arcsec wide bins
placed along the filaments, giving the peaked surface brightness
histogram shown in Fig. 4. This is
consistent with the ICM particle heating model (Fabian et al 2011)
where heating and excitation is the result of the uniform flux of hot
ICM particles, provided that the covering fraction of a filament is
high, despite the low volume filling factor.

The shape of the filaments to the E strongly suggests that they are
being drawn outward by the outflowing hot gas as part of the bubbling
cycle. Consider the bright filament arc 19 arcsec to the ESE of the
nucleus, just outside the Western bubble (Fig. 5), which has radius,
$R\sim 1\kpc$ and thickness $d\sim60\pc$. Assuming that the hot gas,
density $\rho_h$, dragging the arc out at velocity $v=10^2v_2 \kmps$ is balanced
by an internal magnetic field $B$, then
\begin{equation} \frac{1}{2}\frac{\rho_h v^2}{d}=\frac{ B^2}{4\pi R}
\end{equation} 
from which we deduce that $B\sim70v^2_2 \mu G$, which equals the
thermal pressure (Sanders et al 2016), assuming a reasonable value for
the drag velocity 
(Canning et al 2011) of $v_2\sim1.$ $\rho_{\rm h}$ corresponds to the 
X-ray measured value (Sanders et al 2016) at that radius of 0.3 
particles cm$^{-3}$.
We can also consider the magnetic support of a radial filament of
length $\ell$ against tidal force to give a further value (Fabian et al
2008) of $B$,
\begin{equation} 
\rho_c\pi d^2 g\ell= \frac{B^2}{4\pi}
\end{equation}

\begin{figure*}
  \centering
  \includegraphics[width=0.95\textwidth,angle=0]{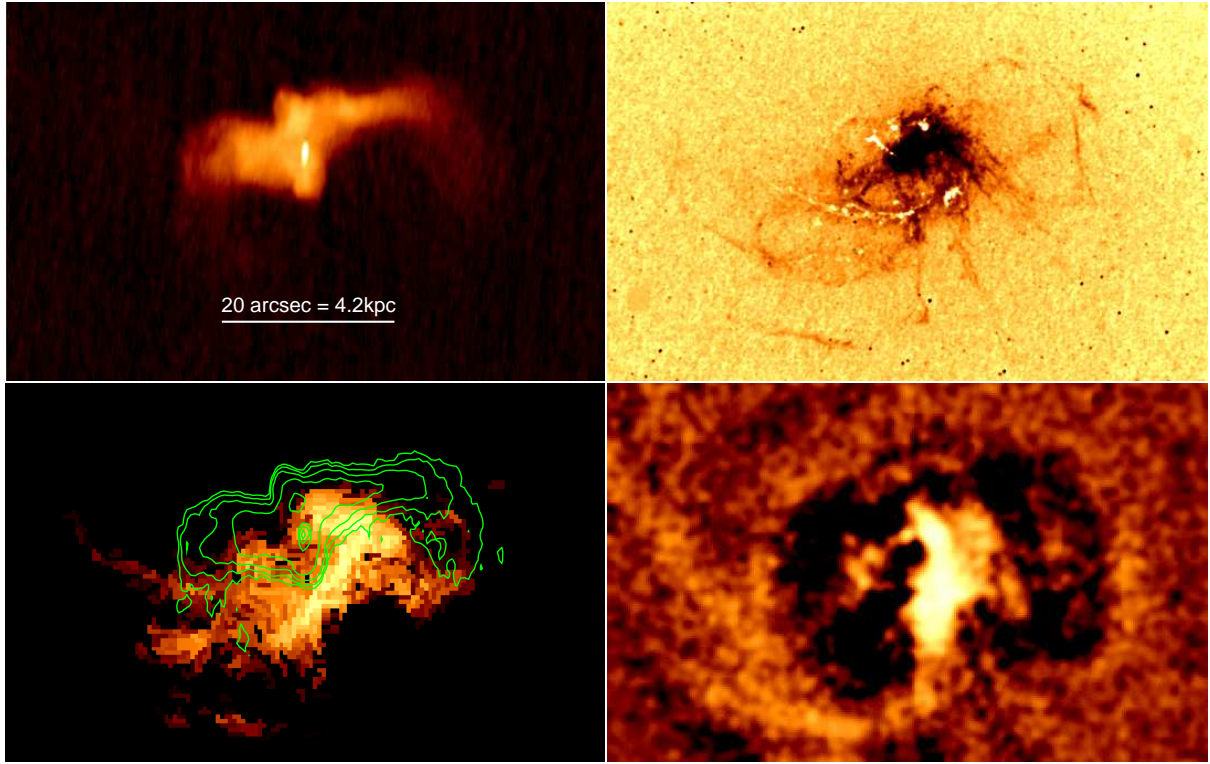}
  \caption{Clockwise from top left: 8 GHz Radio map; H$\alpha$ image; X-ray
emission in the 1-1.5 keV band (adaptively-smoothed) showing the
bubbles as dark patches; soft X-ray 0.5 keV emission (with radio
contours superimposed).  }
\end{figure*}

We obtain $\sim 85 \mu G$ using the value of the gravitational
acceleration $g=1.4\times10^7(r/{\kpc})^{-0.68}\cmpssq$ from
Sanders et al (2016). This value is similar to that found (Fabian et
al 2008) for the filaments in
NGC1275.

The strong Faraday Rotation seen in the radio source (Fig. 5) is
interpreted as due to a mean line-of-sight magnetic field,
$B\sim25\mu G$, in the $5\times10^6 \K$ component of 
the inner hot gas (Taylor et
al 2007). This corresponds to a magnetic pressure about 10 per cent of
the thermal pressure. Filaments are thus magnetised structures
embedded in a less magnetised ICM. Note that if hot gas at fixed
temperature is clumped so that its volume filling factor is $f_{\rm v}$, then
the value of $B$ deduced from Faraday Rotation scales as $B_{\rm
  RM}\propto v^{1/6}$.

We have used the depth of dust absorption in the filaments to make a
rough assessment of their location along the line of sight (Gillmon et
al 2004). The starlight of the galaxy will fill in the absorption if
the filaments are close to the centre. Fig. 6 displays the radius from
the centre of NGC4696 at which the lowest flux in a dust patch matches
the flux in that band from the galaxy. Most of the prominent outer
filaments are then situated 4-5 kpc from the nucleus. We presume that
the filaments without any obvious dust absorption are on the other
side of the galaxy.

\begin{figure}
  \centering
  \includegraphics[width=0.99\columnwidth,angle=0]{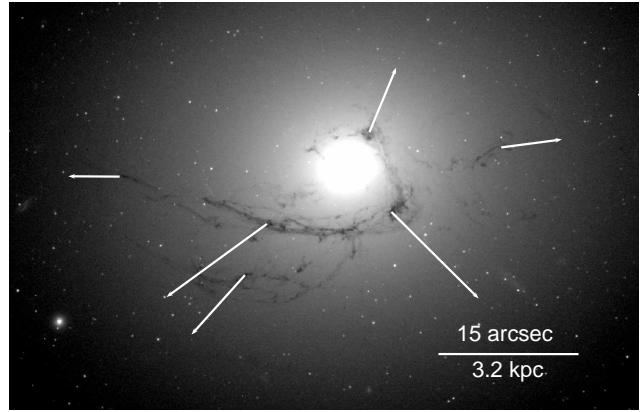}
  \caption{The white vectors on this F467 band image show the distances
from the galaxy centre at which the galaxy surface brightness
decreases to the central surface brightness of 6 regions seen in
absorption. Assuming that the fialments completely block all the
emission from behind them, the vectors give an indication of the
expected length along our line of sight that each filament is
situated.   }
\end{figure}

An abundance drop is seen in Chandra X-ray spectral imaging of the
centre of NGC4696 with iron peaking at $1.8 \Zsun$ at a radius of
15~kpc and dropping inward to $0.4 \Zsun$ in the innermost $4\kpc$
(Panagoulia, Fabian \& Sanders 2013). This can be explained if iron
and many other metals injected by stellar mass-loss in the central
region are locked up in grains which contribute to the filamentary
nebula. The quasi-continuous bubbling process drags the filaments
outward and dumps them beyond 10 kpc where they mix and join the
ICM. An approximate steady state is reached if about 10 per cent of
the dusty filament mass is dragged out every cycle (Panagoulia et al
2013), which corresponds to a mean outflow rate of $\sim 1 \Msunpyr$.

In Fig. 7, we show the XMM RGS spectrum from the inner region of the
cluster core of width 5 kpc. Emission lines from FeXVII,
characteristic of gas at 0.5 keV temperature are seen as well as a 3$\sigma$
detection of the OVII triplet near 22 A from gas at 0.2 keV. The
latter corresponds to a radiative cooling rate of $1.0 \Msunpyr$ over
the temperature range 0.3 to 0.05 keV, assuming Solar abundances. We
have no clear evidence that the gas is cooling but note that the
radiative cooling time of the 2 MK component is only $10^7 \yr$. The
gas may however be part of the mixing interface between the cold
filaments and the hot surrounding medium (Chatzikos et al 2015).

A filament is therefore a quasi-linear aggregation of dense cold knots
having a combined covering factor of unity over a width of about 60
pc, linked together by strong magnetic fields, permeated and
surrounded by a volume-filling hot component at about $5\times10^6
\K$. The hot component is itself magnetized but to a lesser level than
the embedded filaments.

\section{The Nuclear Region} It is interesting to compare the mass of
cold gas with that of the surrounding hot gas. The electron density of
the hot gas in the central 30 kpc is approximately (Sanders et al
2016) $0.168 (r/kpc)^{-0.83} \pcmcu$. The (projected) H$\alpha$ emission
scales as $r^{-0.8}$ so, assuming that the cold gas mass scales
similarly, the region within 0.5 kpc of the nucleus is dominated by
cold gas.

The Bondi radius for accretion, $R_B= 2GM/c_s^2$, onto the central
black hole is 70 pc if we assume that the volume-filling component is
the cooler X-ray emitting gas at $5\times10^6\K$ and use a black hole
mass $M_{\rm bh} \sim10^9\Msun$. The mass of the central black hole in
NGC4696 has not yet been measured directly, but this is the mass
deduced using the measured stellar velocity dispersion (Bernardi et al
2002) $\sigma= 254\kmps$ applied to the $M_{\rm bh} - \sigma$ relation
(McConnell \& Ma 2013). The proximity and likely high mass of the
black hole in NGC4696 mean that its accretion radius is one of very
few in an active system which is currently resolvable.

The HST imaging shows that the H$\alpha$ emission peaks near the
nucleus coincident with a swirling S-shape or total diameter 1 arcsec
(210 pc), which presents a similar aspect to that of the much larger
filament system (Fig. 8). Most of the swirl lies just outside or
within the Bondi radius and presumably indicates that the flow
proceeds as a connected structure, rather than a chaotic rain of cold
blobs (see Gaspari, Brighenti \& Temi 2015) for hydrodynamic
simulations of a blobby inflow). The total mass of gas in the stream,
assuming a constant H$\alpha$ to gas mass ratio, is about
$5\times10^6 \Msun$ which must accrete slowly to match the accretion
rate required by the radio source of about
$10^{-2} \eta_{0.1} \Msunpyr$, where $\eta_{0.1}$ is the radiative
efficiency in units of 0.1. (The free-fall time from the Bondi radius
is about $2\times10^5\yr$.) There is again the old problem (Fabian \&
Canizares 1988) of ``why is
the nucleus of many local massive ellipticals dead?'' when there is
sufficient gas to make them as active as quasars. This time it
involves cold rather than hot gas.

The swirling structure strongly indicates that the gaseous atmosphere
around the Bondi radius is slowly rotating and the filamentary
appearance means that magnetic fields remain important. Angular
momentum may stem the inflow at radii smaller than resolved here. It
is possible that the hot phase in the inner few 100 pc participates in
a giant advection-dominated accretion flow (ADAF: Narayan \& Fabian
2011). Estimating the accretion rate when the gas is highly
multiphase is difficult, but the region is not starved of cold gas,
which forms the dominant component

The H$\alpha$ brightest patch coincides with a radio knot imaged by
the VLBA (Fig. 9; Taylor et al 2006) and could be where the outgoing
jet passes close to the gas stream. A continuation outward along the
jet leads to a FUV source, which could be a synchrotron knot or
hotspot in the jet. The jet structures explain the ``double nucleus''
reported earlier (Laine et al 2003) and the emission attributed to
central star formation Tremblay et al (2015).
\begin{figure}
  \centering
  \includegraphics[width=0.99\columnwidth,angle=0]{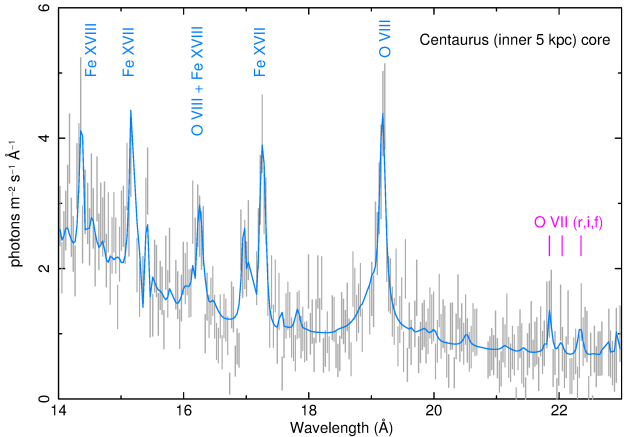}
  \caption{XMM RGS spectrum of the inner 5 kpc region of the cluster core
around the nucleus. The strong FeXVII and Oxygen lines show that the
region is multiphase. OVII is detected at just above 3? significance,
indicating the presence of gas at 2 million K.  }
\end{figure}

\begin{figure}
  \centering
  \includegraphics[width=0.99\columnwidth,angle=0]{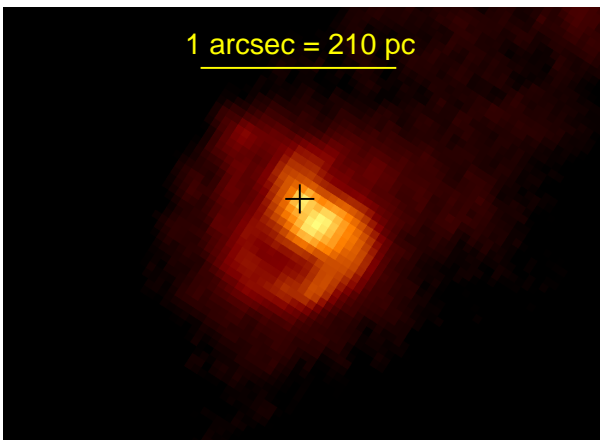}
  \caption{H$\alpha$ emission from the innermost few arcsec around the nucleus
(marked by a cross), revealing a 150pc swirl of gas
   }
\end{figure}

It should be possible to map the velocity of the swirling cold gas
with the full array of ALMA or in the optical/IR band with adaptive
optics or HST spectroscopy.

\begin{figure}
  \centering
  \includegraphics[width=0.99\columnwidth,angle=0]{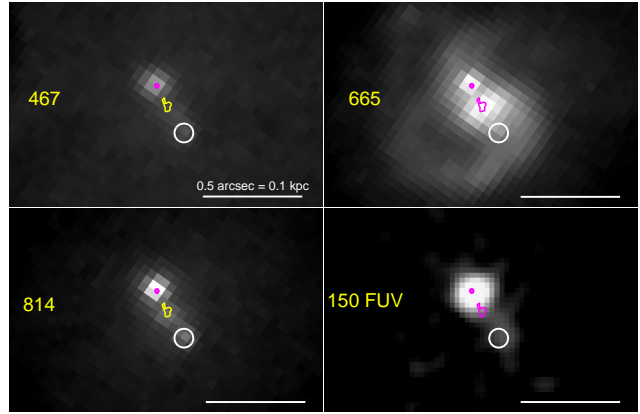}
  \caption{The innermost region imaged in different filters. The
magenta/yellow contours show the Very Long Baseline Array (VLBA) radio
emission with the upper point being the nucleus and the lower patch a
hotspot in the jet. The numbers in the panels refer to the HST filters
used (F467M, F665N, F814W and FUV150).   }
\end{figure}

\section{Discussion} HST imaging has resolved the dusty molecular
filaments in NGC4696 to be about 60 pc in width, slightly less than
the filaments around NGC1275 in the Perseus cluster. The surface
brightness of the outer filaments appears to be constant on a half
arcsec scale. The filaments near the centre appear brighter, possibly
due to the overlap of filaments along our line of sight. Estimates of
the magnetic field required to support the filaments as they are drawn
out with the hot gas and against tidal forces indicate that they are
in rough equipartition with the hot gas. Magnetic fields in the hot
gas compatible with Faraday rotation measures from the extended radio
source are about one third of this value.

The lack of star formation is probably due to the strong internal
magnetic field (Fabian et al 2008). The critical surface density for
gravitational instability in the presence of a magnetic field is given
(Zweibel et al 1993) by $\Sigma_{\rm c} \sim B / (2\pi G^{1/2})$; for $75 \mu$G
this requires $\Sigma > 0.04 \gpcmsq$ for collapse. The transverse
column density in a filament is about 1.5 orders of magnitude lower
than this value, so the filaments should be stable against
gravitational collapse and subsequent star formation.

The hot gas in the central few kpc (Fig. 10) is multiphase with
components detected around 1 keV, 0.5 keV and also 0.2 keV, by XMM
spectroscopy and Chandra imaging. The ratios of the components are not
compatible with a simple cooling flow, although a sporadic flow of a
few $ \Msunpyr$ may occur. There is however no sign of any star
formation.

The HST imaging has resolved the Bondi radius of the central black
hole where we find a large 100 pc radius swirl of gas. The
surroundings of the swirl are also filamentary, suggesting that
magnetic fields are dynamically important in the cold gas at all
radii. The behaviour of the inner gas of this massive elliptical
galaxy, and probably many other such galaxies, is mediated by strong
magnetic fields in both the hot gas and the embedded cold, molecular,
filaments.

\begin{figure}
  \centering
  \includegraphics[width=0.99\columnwidth,angle=0]{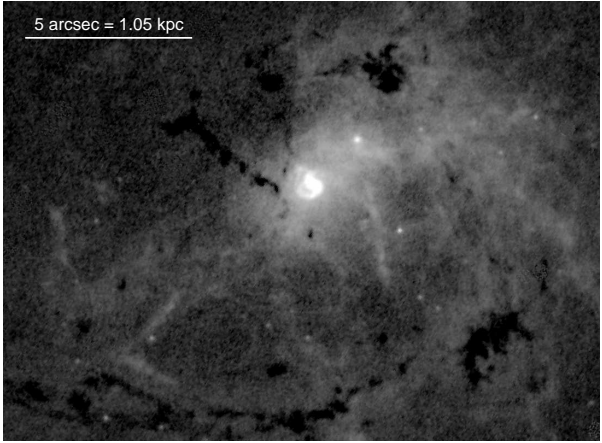}
  \caption{A wider H$\alpha$ image of the nucleus and surrounding gas and
dust. The image is 3.2 kpc across.   }
\end{figure}

\section*{Acknowledgements}
ACF, SAW and HRR acknowledge support from ERC Advanced Grant FEEDBACK,
340442. The work is based on observations made with the NASA/ESA Hubble Space
Telescope, obtained at the Space Telescope Science Institute, which is
operated by the Association of Universities for Research in Astronomy,
Inc., under NASA contract NAS 5-26555. These observations are
associated with program 13821. 



\section{Appendix}

\subsection{HST Data reduction and Analysis}

Fourteen observations of NGC 4696 were taken with HST, and these are
listed in Table 1. We obtained 10.5 orbits using the redshifted
H$\alpha$ filter F665N, 2 orbits using the blue filter F467M, and half
an orbit using the red filter F814W.The observations were all aligned
using DrizzlePac 2.0 (drizzlepac.stsci.edu), which uses the positions
of the stars in the images identified by TweakReg to match their
coordinates. The 11 H$\alpha$  images were then mosaicked together using
astrodrizzle to produce the total H$\alpha$ images. The two red band
images were mosaicked in the same way.

We then removed the galaxy continuum from the H$\alpha$ image to
maximise the contrast of the H$\alpha$ filaments. This was achieved by
scaling a heavily-smoothed red band image (smoothing with a Gaussian
kernel with a FWHM of 40 pixels) and subtracting it from the H$\alpha$
image. Point sources were removed from the red band image before doing
this, and replaced with random pixel values from neighbouring
locations to ensure a smooth shape, matching the galaxy continuum
shape in the H$\alpha$  images to give Fig. 1.

\begin{table*}
\begin{center}
\caption{HST observations used}
\label{obsids}
\leavevmode
\begin{tabular}{ llllllllll } \hline \hline
Dataset & Target Name & RA & Dec& Obs date & Exp time & Filter & Central Wavelength  \\ \hline
ICLE02030 & NGC-4696	& 12:48:49.270	& -41:18:39.01 & 	2015-04-27 & 	2684.000	& F467M	& 4682.594	 \\
ICLE02020 & NGC-4696	& 12:48:49.270	& -41:18:39.01 & 	2015-04-27 & 	1155.000	& F814W	& 8026.814	 \\
ICLE02010 & NGC-4696	& 12:48:49.270	& -41:18:39.01 & 	2015-04-27 & 	1134.000	& F665N	& 6655.850	 \\
ICLE02040 & NGC-4696	& 12:48:49.270	& -41:18:39.01 & 	2015-04-27 & 	2684.000	& F467M	& 4682.594	 \\
ICLE01010 & NGC-4696	& 12:48:49.270	& -41:18:39.01 & 	2015-06-24 & 	2572.000	& F665N	& 6655.850	 \\
ICLE01020 & NGC-4696	& 12:48:49.270	& -41:18:39.01 & 	2015-06-24 & 	2684.000	& F665N	& 6655.850	 \\
ICLE01030 & NGC-4696	& 12:48:49.270	& -41:18:39.01 & 	2015-06-24 & 	2684.000	& F665N	& 6655.850	 \\
ICLE01050 & NGC-4696	& 12:48:49.270	& -41:18:39.01 & 	2015-06-24 & 	2684.000	& F665N	& 6655.850	 \\
ICLE01040 & NGC-4696	& 12:48:49.270	& -41:18:39.01 & 	2015-06-24 & 	2684.000	& F665N	& 6655.850	 \\
ICLE03030 & NGC-4696	& 12:48:49.270	& -41:18:39.01 & 	2015-08-04 & 	2684.000	& F665N	& 6655.850	 \\
ICLE03020 & NGC-4696	& 12:48:49.270	& -41:18:39.01 & 	2015-08-04 & 	2684.000	& F665N	& 6655.850	 \\
ICLE03040 & NGC-4696	& 12:48:49.270	& -41:18:39.01 & 	2015-08-04 & 	2684.000	& F665N	& 6655.850	 \\
ICLE03010 & NGC-4696	& 12:48:49.270	& -41:18:39.01 & 	2015-08-03 & 	2572.000	& F665N	& 6655.850	 \\
ICLE03050 & NGC-4696	& 12:48:49.270	& -41:18:39.01 & 	2015-08-04 & 	2684.000	& F665N	& 6655.850	 \\ \hline
\label{HST_obs}
\end{tabular}
\end{center}
\end{table*}

\subsection{APEX CO}

The observations used 2 hr of allocated time (about half on source)
with an averaged system temperature of 200 K and primary beam size of
27 arcsec. We
estimate a molecular mass $10^8 \Msun$ assuming a luminosity distance of
44.3 Mpc and an CO(2-1)/CO(1-0) ratio of 0.7, using the following fit
parameters (errors): Area  $T_{peak}dV$ 421.6 (123.0)  mK km
s$^{-1}$; Line centre $v_0$ 39.0( 17.8)  km~s$^{-1}$; Width
$dV$  107.7 ( 33.1)~km~s$^{-1}$; $T_{peak}$ 3.68 mK. 

The mass was derived by using a Milky Way conversion factor of $4.3
\Msun$ (K km s$^{-1}$ pc$^{-2})^{-1}$ (Bolatto et al 2013).

\subsection{XMM-Newton/RGS data reduction and modelling}

The XMM-Newton RGS camera consists of two similar detectors, which
have high spectral resolution between 6 and 38 Å. The data for
observation ID=0406200101, which is the longest ($\sim120$~ks) on-axis
pointing of NGC 4696 were reduced with the XMM-Newton Science Analysis
System (SAS) v13.5.0 using the latest calibration files. Contamination
from soft-proton flares was corrected following the XMM-SAS standard
procedures. We processed the RGS data with the SAS task {\sc rgsproc} and
the MOS 1 data with {\sc emproc}, respectively, in order to produce event
files, spectra, and response. The total clean time was about $107$~ks
s. The first-order RGS 1 and 2 spectra were extracted in a
cross-dispersion region of 0.4 arcmin width centred on the emission
peak (corresponding to 5 kpc). This was done within {\sc rgsproc} by setting
the {\sc xpsfincl} mask to include 66 per cent of point-source events inside the
spatial source extraction mask. The model background spectrum was
created by {\sc rgsproc}.

We fit the RGS 8-27 A (0.46-1.55 keV) spectra with the SPEX package;
outside this energy band, the S/N ratio is too low. We bin the RGS
spectra in channels equal to 1/3 of the PSF, and use C-statistics,
because it provides the optimal spectral binning and avoids over-sampling.

The hot gas in NGC 4696 is multiphase and we first fit the RGS spectra
with a combination of two components of gas in collisional ionization
(two {\sc cie} models in SPEX with $T_1 = 1.68\pm0.05 keV$ and $T_2 = 
0.74\pm0.01
\keV$, i.e. $\sim2 × 10^7 \K$ and $9 \times 10^6 \K$, respectively), 
both absorbed by
Galactic absorption ($N_{\rm H} = 8.56 \times 10^{20} \pcmsq)$ and corrected by redshift
($z = 0.010$). The emission-line broadening due to the spatial extent of
the source was corrected by convolving the emission components by the
{\sc lpro} model in SPEX, which takes as input the MOS 1 source spatial
profile.

We search for cooler gas through the O VII line triplet near 22A,
which was recently discovered in a sample of nearby, bright,
elliptical galaxies (Pinto et al 2014). This was done in two
alternative ways: 
at first we add a third {\sc cie} component ($T_3 = 0.13 \pm 0.02 \keV$,
 i.e. $kT\sim1.5 \times
10^6\K$) to fit any line from cool gas including the O VII resonant
(21.6A), intercombination (21.8A), and forbidden (22.1A) lines, also
convolved by the spatial broadening (see Fig. 7). Alternatively, we
tested two delta lines at the position of the O VII stronger (21.6A
and 22.1A) lines, also convolved by the spatial broadening, but
coupling their normalization to the value given by
collisionally-ionized gas (R 21.6A/ 22.1A $\sim3/2$). In both cases we
obtain a detection significance above 3$\sigma$. The O VII lines can also be
fitted with a cooling flow model providing a cooling rate of about $1
\Msunpyr$. In all the spectral fits the abundances of N, O, Ne, Mg, Ca,
Fe, and Ni were free to vary but coupled between the cie components;
they range between 0.4x and 1.3x solar.

\end{document}